\documentclass[prb,letterpaper,twocolumn,aps,floatfix,longbibliography]{revtex4-2}
\usepackage{graphicx}
\usepackage{xcolor}
\usepackage{amsmath}
\usepackage{amsfonts}
\usepackage[small,bf]{subfigure}
\newcommand{\beq}{\begin{equation}}
\newcommand{\eeq}{\end{equation}}
\newcommand{\bk}{{{\bf{k}}}}

\newcommand{\br}{{{\bf{r}}}}

\newcommand{\bq}{{\bf{q}}}

\newcommand{\bn}{{\bf{n}}}

\newcommand{\beqa}{\begin{eqnarray}}
\newcommand{\eeqa}{\end{eqnarray}}
\newcommand{\pdg}{{\vphantom \dag}}
\newcommand{\dg}{{\dag}}
 
\newcommand{\bsigma}{{\boldsymbol \sigma}}
\newcommand{\btau}{{\boldsymbol \tau}}

\newcommand{\ra}{\rightarrow}

\newcommand{\cL}{{\cal L}}
\begin{document}
\title{Disordered Weyl semimetal as an array of coupled Hubbard chains}
\author{Jinmin Yi}
\affiliation{Department of Physics and Astronomy, University of Waterloo, Waterloo, Ontario 
N2L 3G1, Canada} 
\affiliation{Perimeter Institute for Theoretical Physics, Waterloo, Ontario N2L 2Y5, Canada}
\author{A.A. Burkov}
\affiliation{Department of Physics and Astronomy, University of Waterloo, Waterloo, Ontario 
N2L 3G1, Canada} 
\affiliation{Perimeter Institute for Theoretical Physics, Waterloo, Ontario N2L 2Y5, Canada}
\date{\today}
\begin{abstract}
We demonstrate that a disordered magnetic Weyl semimetal may be mapped onto a two-dimensional array of coupled 
replicated Hubbard chains, where the Hubbard $U$ is directly related to the variance of the disorder potential. This is a three-dimensional 
generalization of a similar mapping of the two-dimensional quantum Hall plateau transition to a one-dimensional 
Hubbard chain. We demonstrate that this mapping leads to the conclusion that the Weyl semimetal becomes a diffusive metal with a nonzero 
density of states at arbitrarily weak disorder, in agreement with recent work.  We also discuss the absence of localization in strongly disordered Weyl semimetals from the 
viewpoint of this mapping.
\end{abstract}
\maketitle
\section{Introduction}
\label{sec:1}
The interplay of disorder and nontrivial electronic structure topology is known to produce highly nontrivial phenomena. 
The best example is the integer quantum Hall effect (IQHE), where disorder-induced Anderson localization is essential in creating 
sharply-quantized Hall conductance, which arises due to the presence of edge states, mandated by the nontrivial topology of Landau levels. 
A magnetic Weyl semimetal~\cite{Murakami07,Wan11,Burkov11-1,Burkov11-2,Weyl_RMP} particularly in its simplest incarnation, with only a pair of nodes and no other states at the Fermi energy~\cite{Burkov11-1,Burkov11-2} offers a generalization of the two-dimensional (2D) IQHE physics to three dimensions (3D). 
Such a Weyl semimetal has in fact very recently been achieved experimentally in Cr-doped Bi$_2$Te$_3$~\cite{Belopolski25} which 
makes such a generalization of the quantum Hall physics to 3D a reality. 

The analogy to the IQHE is based on the fact that a magnetic Weyl semimetal with a pair of nodes may be regarded as an intermediate phase between a 3D quantum Hall insulator and an ordinary insulator. The 3D quantum Hall insulator, in turn, may be viewed as a stack of 2D quantum Hall insulators~\cite{Halperin87,Kohmoto92}.
The IQH to ordinary insulator transition (plateau transition) is sharp in 2D, since the Hall conductivity is quantized to integer multiples of $e^2/h$, 
the Hall conductance quantum, in weakly-interacting 2D insulators. 
In 3D, however, the Hall conductivity is no longer dimensionless in units of $e^2/h$, but has units of inverse length, or wavevector. 
This wavevector is equal to a reciprocal lattice vector in the 3D Hall insulator and there is no reason it needs to sharply jump to zero at the transition 
to the ordinary insulator. Instead, it changes continuously, and the 2D plateau transition broadens into a 3D Weyl semimetal phase. 
The Hall conductivity of this Weyl semimetal is continuously-tunable and is proportional to the separation 
between the Weyl nodes in the first Brillouin zone (BZ)~\cite{Burkov11-1,Burkov11-2,Burkov14-2}.

This analogy of the 3D magnetic Weyl phase to the 2D plateau transition suggests that at least some of the phenomena, observed in 2D quantum Hall systems, 
may also have 3D analogs in Weyl semimetals. Indeed, it has been demonstrated that 3D analogs of fractional quantum Hall effect are possible, when strong 
electron-electron interactions are included~\cite{Wang20,Yi23}.
In addition, in an exact analogy to the 2D plateau transition, lack of localization in magnetic Weyl semimetals 
(assuming the 3D quantum Hall insulator phase is not destroyed by disorder) has been demonstrated~\cite{Yi24}.

In this paper we further explore the 2D plateau transition analogy in order to understand the effects of weak disorder in Weyl semimetals. 
While it is agreed that strong enough disorder destroys the Weyl nodes and creates a nonzero density of states, turning a Weyl semimetal 
into a diffusive metal (which still has nontrivial topological properties, inherited from the clean Weyl semimetal), what happens at weak disorder has been 
controversial~\cite{Arovas13,Brouwer14,Nandkishore14,Syzranov15,Wang15,Altland15,Pixley15,Xie15,Altland16,Hughes16,Pixley16,Brouwer16,Khalaf17,Syzranov18,Altland18,Mirlin19,Alisultanov24}.
Both scaling arguments~\cite{Burkov11-2} and direct self-consistent Born approximation (SCBA) calculations~\cite{Fradkin86-1,Fradkin86-2,Brouwer14,Syzranov15} suggest a phase transition from a ballistic Weyl semimetal to a diffusive metal at a finite disorder strength. 
However, it has been argued~\cite{Nandkishore14,Pixley15,Pixley16} that these considerations miss the effect of rare regions, i.e. rare strong impurity potential fluctuations, which always create a nonzero, albeit exponentially small at weak disorder, density of states, which leads to diffusive transport. 

Here we will use an exact mapping from a disordered magnetic Weyl semimetal with a pair of nodes onto a 2D array of coupled replicated Hubbard chains (in the limit when the number of replicas $R$ is taken to zero), where 
the Hubbard $U$ is directly related to the variance of the disorder potential, to shed new light on this issue. 
This mapping is a direct generalization of the well-known mapping of the 2D IQHE plateau transition onto a 1D Hubbard chain~\cite{Lee94,Lee96,Kim96,Balents97}.
Within this picture the IQH and ordinary insulators map onto band insulators without magnetic order, which arise due to strong enough dimerization, breaking translational symmetry in the Hubbard chains. The Weyl phase maps onto either a gapless 2D Dirac liquid (ballistic Weyl semimetal) or a Mott insulator with antiferromagnetic order (diffusive metal).
We demonstrate that the former never happens and the Weyl phase is in fact a diffusive metal at any nonzero disorder strength. Moreover, we argue that 
the nonperturbative effect of rare regions translates into similarly nonperturbative Mott physics in the language of the Hubbard chain array. This contrasts with perturbative
generation of a finite density of states within SCBA, corresponding to the Hartree-Fock magnetic instability at a finite interaction strength in the Hubbard chains array language. 

The rest of the paper is organized as follows. 
In Section~\ref{sec:2} we describe the exact mapping between a magnetic Weyl semimetal with a pair of nodes in a gaussian-distributed impurity potential 
and an array of coupled Hubbard chains at half-filling. The Hubbard $U$ is directly related to the variance of the disorder potential. 
In the $U = 0$ limit, the band structure of the array contains a pair of 2D Dirac points, which are the remnants of the Weyl points of the original Weyl semimetal. 
In Section~\ref{sec:3.1} we first analyze the Hubbard array in the Hartree-Fock (HF) approximation, which is equivalent to SCBA, and show that 
the gapless 2D Dirac phase is stable against gap opening and the development of magnetic order up to a finite critical value of $U$, as expected. 
In Section~\ref{sec:3.2} we demonstrate that the HF analysis misses Mott transition physics and, in the limit $R \ra 0$, the array is in fact always in the Mott phase. 
Furthermore, we argue that there is always a magnetic order  in the Mott phase, corresponding to a diffusive metal in the original 3D Weyl system. 
In Section~\ref{sec:4} we discuss the absence of localization in a strongly disordered Weyl semimetal from the viewpoint of the coupled Hubbard chain mapping 
and rederive the nonlinear sigma model (NLSM) with a topological term, which describes the disordered Weyl metal and which may be viewed as a 3D generalization 
of the Pruisken's NLSM~\cite{Pruisken90}.
We conclude with a brief summary and a discussion of the results in Section~\ref{sec:5}. 

\section{Mapping a magnetic Weyl semimetal to a Hubbard chain array}
\label{sec:2}
We start from the model of magnetic Weyl semimetal with a pair of nodes, introduced in Ref.~\cite{Burkov11-1}. 
This is the simplest model of a Weyl semimetal, which nevertheless does describe an experimentally-realized system~\cite{Belopolski25}.
The Hamiltonian of the multilayer is given by
\beqa
\label{eq:1}
H&=&\sum_{\bk_{\perp}, z, z'} c^\dg(\bk_{\perp}, z) \left[-v_F \tau^z (\hat z \times \bsigma) \cdot \bk_{\perp} \delta_{z, z'}  \right. \nonumber \\
&+& \left. (b \sigma^z + t_S \tau^x) \delta_{z, z'} + \frac{1}{2} t_D \tau^+ \delta_{z', z+1}\right. \nonumber \\
&+&\left.\frac{1}{2} t_D \tau^- \delta_{z', z-1} \right] c^\pdg(\bk_{\perp}, z'). 
\eeqa
This Hamiltonian describes a 3D magnetic topological insulator (TI), such as the Cr-doped Bi$_2$Te$_3$ of Ref.~\cite{Belopolski25}.
As any TI, its electronic structure may be modelled as a layered structure, consisting of its surface states (2D Dirac fermions in our case), which are 
coupled by tunneling matrix elements. Each unit cell of the structure contains a pair (labeled by the eigenvalues of the Pauli $\tau^z$ operator) 
of 2D Dirac fermions, related by inversion symmetry. $t_{S,D}$ are tunneling matrix elements, coupling the 2D Dirac states in the same (S) or neighboring (D)
unit cells. Further-neighbor tunnelings are in principle always present, but are ignored in this model. 
This makes the electronic structure particle-hole symmetric. While this symmetry is not an exact symmetry of any real Weyl semimetal, it should not affect the 
physical properties we are interested in and it will prove very convenient to explicitly keep this symmetry, as will be seen below. 
$v_F$ is the Fermi velocity, characterizing the 2D Dirac states and $\bk_{\perp}$ is the crystal momentum in the $xy$-plane ($\hbar = c = e =1$ units will be used throughout). 
$\bsigma$ is the triplet of Pauli matrices, describing real spin and $b$ is the exchange spin splitting, produced by the ferromagnetic order. 
Both spin $\bsigma$ and pseudospin $\btau$ indices are kept implicit in Eq.~\eqref{eq:1} in order not to clutter the notation. 
Finally, $z, z'$ are discrete coordinates of the unit cells, stacked in the $z$-direction.

We now want to modify Eq.~\eqref{eq:1} to a form that will be suitable for the mapping to an array of Hubbard chains. 
We will start by reinterpreting our system as a stack of coupled 2D quantum Hall insulators. 
To reveal this, we first make a transformation
\beq
\label{eq:2}
\sigma^{\pm} \ra \tau^z \sigma^{\pm}, \,\, \tau^{\pm} \ra \sigma^z \tau^{\pm}, 
\eeq
which brings Eq.~\eqref{eq:1} to the form
\beqa
\label{eq:3}
H&=&\sum_{\bk_{\perp}, z, z'} c^\dg(\bk_{\perp}, z) \left[- v_F (\hat z \times \bsigma) \cdot \bk \delta_{z, z'}  \right. \nonumber \\
&+& \left. (b + t_S \tau^x) \sigma^z \delta_{z, z'} + \frac{1}{2} t_D \tau^+ \sigma^z \delta_{z', z+1}\right. \nonumber \\
&+&\left.\frac{1}{2} t_D \tau^- \sigma^z \delta_{z', z-1} \right] c^\pdg(\bk_{\perp}, z'). 
\eeqa
Fourier transforming the discrete $z$ coordinates and partially diagonalizing the Hamiltonian in the pseudospin subspace, we obtain 
the following momentum-space Hamiltonian
\beq
\label{eq:4}
H_r(\bk) = - v_F (\hat z \times \bsigma) \cdot \bk + m_r(k_z) \sigma^z, 
\eeq
where $ r = \pm 1$ and 
\beq
\label{eq:5}
m_r(k_z) = b + r \sqrt{t_S^2 + t_D^2 + 2 t_S t_D \cos(k_z)}. 
\eeq
This has the form of a 2D Dirac Hamiltonian with a mass that depends on $k_z$. Taking $b$ to be positive, the $m_+(k_z)$ mass is always 
positive, while $m_-(k_z)$ changes sign at $k_z = \pi \pm Q$, where 
\beq
\label{eq:6}
Q = \arccos \left(\frac{t_S^2 + t_D^2 - b^2}{2 t_S t_D} \right), 
\eeq
which are the BZ locations of a pair of Weyl nodes. The nodes exist as long the exchange spin splitting is in the interval $|t_S - t_D| < b < t_S + t_D$.  

We may now eliminate unnecessary extra degrees of freedom by projecting onto the low-energy $r = -1$ subspace. 
It is also useful to reinterpret the $H_-(\bk)$ Hamiltonian as the Hamiltonian of a stack of weakly-coupled Chern insulators. 
To do this, we assume $t_D/t_S \ll 1$ and expand to leading order in this small parameter. 
This gives
\beq
\label{eq:7}
H(\bk) = - v_F (\hat z \times \bsigma) \cdot \bk + (b - t_S) \sigma^z - t_D \cos(k_z) \sigma^z, 
\eeq
where we have dropped the $-$ subscript. 
When $t_D = 0$ Eq.~\eqref{eq:7} describes a stack of independent $2D$ layers, which undergo a plateau transition 
from a quantum Hall insulator with $\sigma_{xy} = e^2/h$ when $b > t_S$ to an ordinary insulator with $\sigma_{xy} = 0$ when $b < t_S$. 
The plateau transition is described by a sign change of the mass of a 2D Dirac fermion~\cite{Haldane88,Ludwig94}.
At a finite $t_D$ this plateau transition broadens into a Weyl semimetal phase, which exists in the interval $t_S - t_D < b < t_S + t_D$. 

\begin{figure}[t]
\subfigure{
\includegraphics[width=\columnwidth]{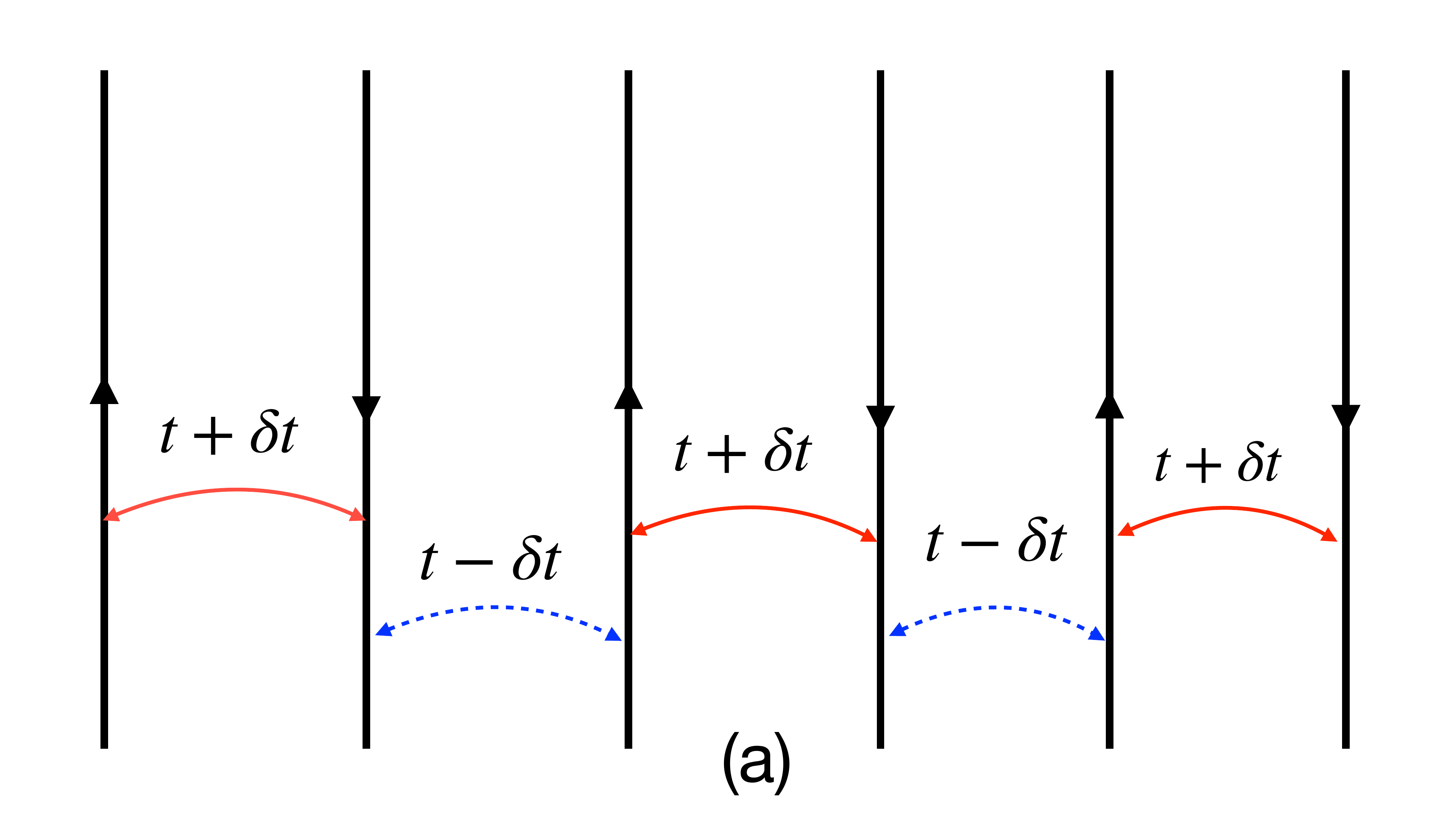}}
\subfigure{
\includegraphics[width=\columnwidth]{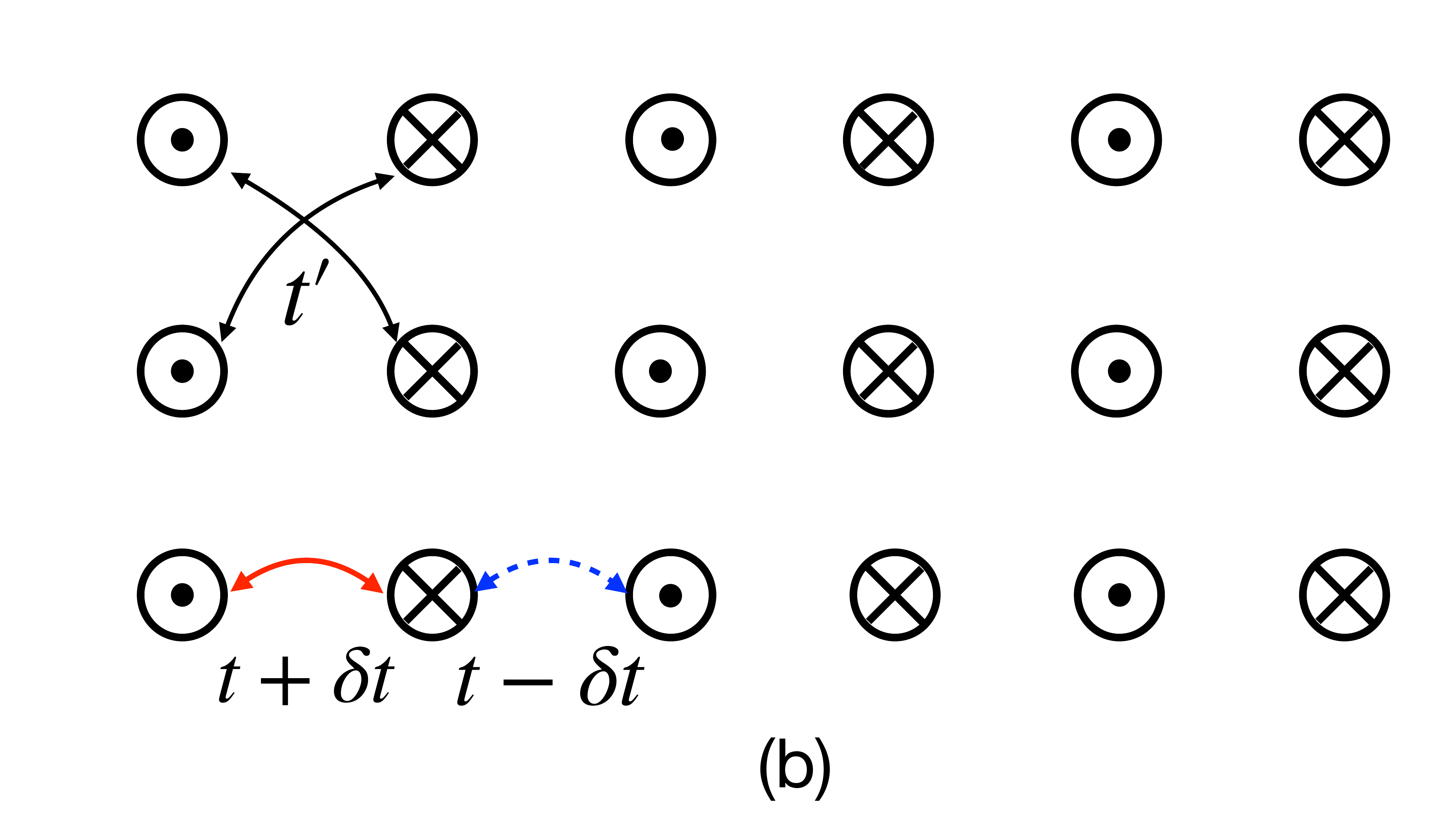}}
\caption{(Color online) (a) Coupled opposite-chirality one-dimensional modes, representing every $xy$-plane of the Weyl semimetal model Eq.~\eqref{eq:9}.
(b) Same array in the $xz$-plane, specifying how opposite-chirality modes are coupled in the $z$-direction in Eq.~\eqref{eq:9}.}
\label{fig:1}
\end{figure}

Now let us rotate the spin quantization axis by $\pi/2$ around the $y$-axis, which transforms $\sigma^z \ra \sigma^x$ and $\sigma^x \ra - \sigma^z$ and gives
\beq
\label{eq:8}
H(\bk) = v_F(\sigma^y k_x + \sigma^z k_y) + [b - t_S - t_D \cos(k_z)] \sigma^x.
\eeq
After this transformation we may reinterpret Eq.~\eqref{eq:8} as describing a network of one-dimensional chiral modes of alternating chirality, running in the $y$-direction,
$\sigma^z$ being the chirality operator~\cite{Lee94}.
Returning to real-space second-quantized notation, we have 
\beqa
\label{eq:9}
H&=&\int d y \sum_{i, j} c^\dg_i(y) \left\{[- i \sigma^z \partial_y - (t + \delta t) \sigma^x] \delta_{i, j} \right. \nonumber \\
&-&\left. \frac{1}{2} (t - \delta t) \sigma^+ \delta_{j, i+ \hat x}  - \frac{1}{2} (t - \delta t) \sigma^- \delta_{j, i - \hat x} \right. \nonumber \\
&-& \left. t' \sigma^x (\delta_{j, i + \hat z} + \delta_{j, i - \hat z}) \right\} c^\pdg_{j}(y). 
\eeqa
Here we have discretized the $x$-direction, while keeping the $y$-direction continuous and removed $v_F$ from the first term by rescaling the $y$ coordinate. 
The indices $i, j = (x, z)$ combine the $x, z$ coordinates and $\hat x, \hat z$ denote primitive translation vectors in the corresponding directions. 
It can be easily seen by Fourier transforming the discrete $x$ and $z$ coordinates, that Eq.~\eqref{eq:9} maps onto Eq.~\eqref{eq:8} with $t = v_F$, 
$\delta t = - (b - t_S)/2$, $t' = t_D/2$ and Weyl points on the $(k_x = \pi, k_y = 0)$ axis. 
The discrete $x$ coordinate labels unit cells in the $x$-direction, each containing a pair of chiral modes, with chirality given by the eigenvalues of $\sigma^z$. 
There is nearest-neighbor tunneling between opposite-chirality modes in the $x$-direction with strength $t + \delta t$, if the two modes belong to the same unit cell, 
and $t - \delta t$, if they belong to neighboring unit cells (i.e. this looks like a Su-Schrieffer-Heeger (SSH) chain of chiral modes), see Fig.~\ref{fig:1}. 
Tunneling in the $z$-direction $t'$ couples opposite-chirality modes within the same $x$-direction unit cell.  
This model may be viewed as a 3D generalization of the Chalker-Coddington network model of the 2D quantum Hall plateau transition~\cite{Chalker88,Chalker95}.

Note that Eq.~\eqref{eq:9} possesses particle-hole symmetry
\beq
\label{eq:10}
c^\dg_i (y) \ra \sigma^z c\pdg_i(y).
\eeq
As discussed above, this is not an exact physical symmetry and is violated, for example, by next-nearest-neighbor tunneling terms, which we neglected. 
Nevertheless, it will prove useful to assume this symmetry is present in what follows. 

As we are interested in a disordered Weyl semimetal, we now add impurity potential to Eq.~\eqref{eq:9}
\beqa
\label{eq:11}
H&=&\int d y \sum_{i, j} c^\dg_i (y) \left\{[- i \sigma^z \partial_y - (t + \delta t) \sigma^x] \delta_{i, j}\right. \nonumber \\
&-&\left. \frac{1}{2} (t - \delta t) \sigma^+ \delta_{j, i + \hat x} - \frac{1}{2} (t - \delta t) \sigma^- \delta_{j, i - \hat x} \right. \nonumber \\
&-& \left. t' \sigma^x (\delta_{j, i + \hat z} + \delta_{j, i - \hat z}) + V_i (y) \delta_{i, j} \right\} c^\pdg_j (y), 
\eeqa
where $V_{i \sigma}(y)$ ($\sigma$ labels the eigenvalues of $\sigma^z$) is a gaussian-distributed 
\beq
\label{eq:12}
P(V) = e^{- \frac{1}{4 U} \int d y \sum_{i \sigma} V_{i \sigma}(y)^2}, 
\eeq  
random potential with zero mean and variance $2 U$
\beq
\label{eq:13}
\langle V_{i \sigma}(y) V_{j \sigma'}(y') \rangle = 2 U \delta_{i, j} \delta_{\sigma, \sigma'} \delta(y - y'), 
\eeq
where the angular brackets denote disorder averaging. 
Note that, after averaging, such a potential does not violate particle-hole symmetry. 
 
At this point let us consider replicated generating functional for the retarded/advanced Green's functions
\beq
\label{eq:14}
Z^R = \int D[\bar c, c] e^{- S[\bar c, c]}, 
\eeq
where $\bar c, c$ are Grassman variables and the action $S$ is given by
\beq
\label{eq:15}
S[\bar c, c] = - \int d y \sum_{i, \sigma, p, a} i \eta p \, \bar c_{i \sigma p a}(y) c_{i \sigma p a}(y) + H(\bar c, c). 
\eeq
Here the index $a = 1, \ldots, R$ labels replicas, $p = \pm 1$ labels the retarded/advanced fields and $\eta = 0+$. 
Differentiating the generating functional with respect to $\eta$ gives the difference between the retarded and advanced Green's functions, 
i.e. the density of states. 
As standard in the replica theory, disorder averaging is evaluated as
\beq
\label{eq:16}
\langle \ln Z \rangle = \lim_{R \ra 0} \frac{1}{R} \langle Z^R - 1 \rangle,  
\eeq 
where $\langle \ldots \rangle$ denotes both disorder and quantum averaging.

Treating $\bar c$ and $c$ as independent fields, we now make a transformation
\beq
\label{eq:17}
c_1 \ra c_1, \,\, \bar c_1 \ra i \bar c_1, \,\, c_2 \ra -i c_2, \,\, \bar c_2 \ra \bar c_2,  
\eeq
where $1,2$ label the eigenvalues of the chirality operator $\sigma^z$. 
This changes the action as
\beqa
\label{eq:18}
&&S[\bar c, c] = \int d y \sum_{i, j} \bar c_i(y) \left\{[\eta p \sigma^z  + \partial_y \right. \nonumber \\
&+& \left. i \sigma^z V_i (y) - (t + \delta t) \sigma^x] \delta_{i, j} \right. \nonumber \\
&-&\left. \frac{1}{2} (t - \delta t) \sigma^+ \delta_{j, i + \hat x}  - \frac{1}{2} (t - \delta t) \sigma^- \delta_{j, i - \hat x}\right. \nonumber \\
&-& \left. t' \sigma^x (\delta_{j, i + \hat z} + \delta_{j, i - \hat z}) \right\} c_j(y)
\eeqa
where we have made the chirality, replica and retarded/advanced indices implicit for brevity. 

Now we reinterpret the $y$-coordinate as imaginary time $y \ra \tau$ and average over the gaussian-distributed disorder potential. 
This gives 
\beqa
\label{eq:19}
&&S[\bar c, c] = \int d \tau \sum_{i, j} \bar c_i(\tau) \left\{[\partial_{\tau} + \eta \sigma^z \tau^z \right. \nonumber\\
&-&\left.(t + \delta t) \sigma^x] \delta_{i, j} - \frac{1}{2} (t - \delta t) \sigma^+ \delta_{j, i + \hat x} \right. \nonumber \\
&-&\left. \frac{1}{2} (t - \delta t) \sigma^- \delta_{j, i - \hat x}  - t' \sigma^x (\delta_{j, i + \hat z} + \delta_{j, i - \hat z}) \right\} c_j(\tau) \nonumber \\
&+&\int d \tau \sum_i  U[\bar c_i(\tau) c_i(\tau)^2], 
\eeqa
where $\tau^z$ is a Pauli matrix in the retarded/advanced pseudospin space. 
This is an imaginary-time action of a 2D repulsive anisotropic Hubbard model.
The infinitesimal positive number $\eta$ now couples to the ``staggered magnetization" $\sigma^z \tau^z$. 
This means that spontaneous magnetic order in this 2D Hubbard model is equivalent to having a nonzero density 
of states at the Fermi energy in the original 3D Weyl semimetal system. 

Passing from imaginary-time action to the Hamiltonian and ignoring infinitesimal 
staggered ``magnetic field" term $\eta \sigma^z \tau^z$, we finally obtain
\beqa
\label{eq:20}
&&H = \sum_{i, j} c^\dg_i \left[- (t + \delta t) \sigma^x \delta_{i, j} - \frac{1}{2} (t - \delta t) \sigma^+ \delta_{j, i + \hat x} \right. \nonumber \\
&-&\left. \frac{1}{2} (t - \delta t) \sigma^- \delta_{j, i - \hat x}  - t' \sigma^x (\delta_{j, i + \hat z} + \delta_{j, i - \hat z})\right] c^\pdg_j \nonumber \\
&+&U \sum_i c^\dg_{i \sigma \alpha} c^\dg_{i \sigma \beta} c^\pdg_{i \sigma \beta} c^\pdg_{i \sigma \alpha}, 
\eeqa
where $\alpha = (p, a)$ is the combined replica and retarded/advanced pseudospin index and we have made summation over all indices, except the spatial 
coordinates, implicit for brevity. 
This completes the mapping of the disordered 3D magnetic Weyl semimetal to a 2D array of coupled Hubbard chains. 
Note that, importantly, this mapping only exists in the case when the particle-hole symmetry is present. 
This is because either next-nearest-neighbor tunneling terms, or a nonzero chemical potential (which would have to be added in their presence 
to keep the filling fixed), would map onto imaginary terms under the transformation of Eq.~\eqref{eq:17}.
\section{Theory of the coupled Hubbard chain array}
\label{sec:3}
\subsection{Hartree-Fock theory}
\label{sec:3.1}
We now analyze the phase diagram of the coupled Hubbard chain theory Eq.~\eqref{eq:20}.
We will first use the simplest Hartree-Fock (HF) approximation, which is equivalent to SCBA in the standard diagrammatic language. 

Decoupling the interaction term using the standard HF procedure we find that the Hartree term vanishes in the replica $R \ra 0$ limit, since it involves two 
independent sums over replica indices. 
This guarantees that even with the Hubbard interaction present, we stay at half-filling in the limit $R \ra 0$. This is an important property that we will use later, even outside of the HF considerations. 
The Fock term gives 
\beq
\label{eq:21}
H_{int} = -U \sum_i \left(2 \langle c^\dg_{i \sigma \alpha} c^\pdg_{i \sigma \alpha} \rangle c^\dg_{i \sigma \alpha} c^\pdg_{i \sigma \alpha} - 
\langle c^\dg_{i \sigma \alpha} c^\pdg_{i \sigma \alpha} \rangle^2 \right). 
\eeq
We define the staggered magnetization order parameter $m$ as
\beq
\label{eq:22}
m = U \langle c^\dg_i \sigma^z \tau^z c^\pdg_i \rangle, 
\eeq
where $\tau^z$ is a Pauli matrix in the retarded/advanced pseudospin space. 
Note that the magnetization is not staggered in the $z$-direction, due to the $z$-direction tunneling pattern shown in Fig.~\ref{fig:1}.
The interaction Hamiltonian may then be written as 
\beq
\label{eq:23}
H_{int} = - m \sum_i c^\dg_i \sigma^z \tau^z c^\pdg_i  + \frac{R N m^2}{U}, 
\eeq
where $N$ is the total number of unit cells. 

The full HF Hamiltonian now takes the form
\beq
\label{eq:24}
H = \sum_{\bk} c^\dg(\bk) H(\bk) c^\pdg(\bk) + \frac{R N m^2}{U},
\eeq
where $H(\bk)$ is an $4 \times 4$ matrix, given by
\beqa
\label{eq:25}
H(\bk) = \left(
\begin{array}{cccc}
- m \tau^z& d_-(\bk) \\
d_+(\bk) & m \tau^z 
\end{array}
\right), \nonumber \\
\eeqa
and
\beqa
\label{eq:26}
d_x(\bk)&=&- (t + \delta t) - (t - \delta t) \cos(k_x) - 2 t' \cos(k_z), \nonumber \\
d_y(\bk)&=&- (t - \delta t) \sin(k_x). 
\eeqa
Note that, in the absence of the $m \sigma^z \tau^z$ term, the Hamiltonian Eq.~\eqref{eq:25} possesses a continuous $SU(2)$ pseudospin 
rotational symmetry, as well as a discrete mirror symmetry
\beq
\label{eq:26.5}
\sigma^x H(-k_x, k_z) \sigma^x = H(k_x, k_z), 
\eeq
both of which are violated by the HF order parameter. 

Diagonalizing Eq.~\eqref{eq:25}, we obtain the HF energy eigenvalues
\beq
\label{eq:27}
\epsilon_{s}(\bk) \equiv s \epsilon(\bk) = s \sqrt{d^2(\bk) + m^2}, 
\eeq
where $s = \pm 1$ label the $2R$-degenerate eigenvalues and 
\beq
\label{eq:28}
d^2(\bk) = d_x^2(\bk) + d_y^2(\bk). 
\eeq
The HF Hamiltonian becomes
\beq
\label{eq:29}
H = \sum_{\bk} \epsilon_{s}(\bk) c^\dg_{s \alpha}(\bk) c^\pdg_{s \alpha}(\bk) + \frac{R N m^2}{U}. 
\eeq
This gives the replicated HF partition function 
\beq
\label{eq:30}
Z^R = \exp\left[-\frac{R N m^2}{U T} + 2 R \sum_{\bk, s}  \ln \left(1 + e^{- \frac{\epsilon_{s}(\bk)}{T}} \right)\right],
\eeq
from which we can obtain the HF free energy
\beqa
\label{eq:31}
F&=&- T \lim_{R \ra 0} \frac{1}{R} (Z^R - 1) \nonumber \\
&=&\frac{N m^2}{U} - 2 T \sum_{\bk, s} \ln \left(1 + e^{- \frac{\epsilon_{s}(\bk )}{T}}\right). 
\eeqa
Minimizing $F$ with respect to $m$ and taking the $T = 0$ limit, we obtain the following HF equation
\beq
\label{eq:35}
1= \frac{U}{N} \sum_{\bk} \frac{1}{\epsilon(\bk)} = \frac{U}{2} \int \frac{d^2 k}{(2 \pi)^2} \frac{1}{\epsilon(\bk)}. 
\eeq
The HF band dispersion Eq.~\eqref{eq:27} contains a pair of 2D Dirac fermions at low energies, which are massless at $m = 0$ and are 
located at 
\beq
\label{eq:36}
k_z = \arccos(- \delta t/ t'), 
\eeq
and $k_x = \pi$. 
These correspond to the locations of the Weyl points in the original 3D model Eq.~\eqref{eq:9}. 
The magnetic order parameter $m$ acts as a mass for the pair of Dirac fermions. 
Since the density of states at the 2D Dirac points vanishes as $\epsilon$ in the vicinity of the Dirac points, it is clear that the 
momentum integral in Eq.~\eqref{eq:35} does not diverge as $m \ra 0$, which means that the HF equation has a nontrivial solution only 
above a finite critical value of $U$. 
This is to be expected, since the HF theory is identical to the SCBA, as already mentioned above. 
\subsection{Slave boson theory of the Mott transition}
\label{sec:3.2}
The HF magnetic ordering transition, described in the previous subsection, is only one possible scenario of magnetic ordering and opening 
of a charge gap in the Hubbard model of Eq.~\eqref{eq:20}. 
The other scenario is a Mott transition, in which a charge gap opens, possibly followed by a subsequent magnetic ordering. 
This is a nonperturbative scenario, that we argue corresponds to the effect of rare regions in the original problem. 

The Mott transition may be described using the standard slave boson approach~\cite{Nagaosa92,Georges_SR,Lee05,Senthil08}.
We represent the electron annihilation operator in Eq.~\eqref{eq:20} as
\beq
\label{eq:37}
c_i = e^{i \theta_i}  f_i. 
\eeq
Here $e^{i \theta_i}$ represents a boson, which carries only the charge of the electron (chargon), while $f_i$ is a $2 R$-component neutral fermion (spinon), which carries the retarded/advanced pseudospin and replica degrees of freedom. For notational convenience, the index $i$ now includes both the unit cell coordinate and the 
chirality index $\sigma$, which may be viewed as an intra-unit-cell coordinate. 
The physical states satisfy the constraint
\beq
\label{eq:38}
n_i = f^\dg_i f^\pdg_i,
\eeq
where $n_{i}$ is the boson number operator, conjugate to the phase $\theta_i$. 
As we have already established above, the Hubbard model Eq.~\eqref{eq:20} is at half-filling in the limit $R \ra 0$. 
Since this is the limit we are interested in, we will take it to be at half-filling at all values of $R$. 

We now decouple the spinons and chargons on each bond in Eq.~\eqref{eq:20} as
\beqa
\label{eq:39}
e^{- i (\theta_i - \theta_j)} f^\dg_i f^\pdg_j &\approx& e^{- i (\theta_i - \theta_j)} \langle f^\dg_i f^\pdg_j \rangle + \langle e^{- i (\theta_i - \theta_j)} \rangle f^\dg_i f^\pdg_j 
\nonumber \\
&-&\langle e^{- i (\theta_i - \theta_j)} \rangle \langle f^\dg_i f^\pdg_j \rangle, 
\eeqa
and then promote the expectation values to fluctuating fields 
\beq
\label{eq:40}
\langle e^{- i (\theta_i - \theta_{i + \mu})} \rangle \ra \Theta_{i \mu} e^{- i a_{i \mu}}, \,\, \langle f^\dg_i f^\pdg_{i + \mu} \rangle \ra \Phi_{i \mu} e^{i a_{i \mu}}, 
\eeq
where $\Theta_{i \mu}$ and $\Phi_{i \mu}$ are real amplitudes, while the phase factors $e^{\pm i a_{i \mu}}$ give rise to compact $U(1)$ gauge fields 
$a_{i \mu}$.~\cite{Barkeshli12}
The time component $a_{i 0}$ accounts for the constraint Eq.~\eqref{eq:38}. 
We will avoid writing out the kinetic terms for the chargons and spinons explicitly due to the unwieldy nature of the kinetic energy term in Eq.~\eqref{eq:20}. 
It is easy to check that this simple decoupling procedure gives identical results to a more complicated generalized Hubbard-Stratonovich transformation 
approach of Ref.~\cite{Lee05}. 

The Hubbard interaction may be written entirely in terms of the chargon number operators as
\beq
\label{eq:41}
H_{int} = U \sum_i (n_i   - R)^2, 
\eeq
which reflects that the Hubbard model is taken to be at half-filling, as discussed above. 

It is clear from the definition that, at the saddle-point level, $\Phi_{i \mu} \sim R$.
It follows that, in the limit $R \ra 0$, the chargons are always in the Mott phase for any value of $U$.
The bosons can then be integrated out, leading to a Maxwell Lagrangian for the gauge field $a_{\mu}$
\beq
\label{eq:42}
\cL_a = \frac{1}{q^2} (\epsilon_{\mu \nu \lambda} \partial_{\nu} a_{\lambda})^2, 
\eeq
where $q^2$ is of the order of the Mott gap, which in turn is of the order of $U$. 

Since, as discussed in Section~\ref{sec:3.1}, the fermion dispersion contains a pair of 2D Dirac band-touching points in the BZ, the low-energy theory then becomes a compact QED$_3$ with $4 R$ flavors of Dirac fermions
\beq
\label{eq:43}
\cL = \bar \psi_a \gamma_{\mu} (\partial_{\mu} + i a_{\mu}) \psi_a + \frac{1}{q^2} (\epsilon_{\mu \nu \lambda} \partial_{\nu} a_{\lambda})^2,
\eeq
where $a$ is the replica index and $\gamma_{\mu}$ are $4 \times 4$ gamma matrices. It is understood implicitly that the gauge field $a_{\mu}$ is compact, so 
monopoles are allowed. 

While for a sufficiently large number of fermion flavors the QED$_3$ may be in a deconfined phase with fermions remaining massless~\cite{Hermele04,Lee08},
as $R \ra 0$ we expect confinement and dynamical mass generation for the fermions~\cite{Herbut03,Seradjeh03,Song19}.
The simplest and most natural mass term is in fact the mass term that arises in the HF theory of magnetic ordering, Eq.~\eqref{eq:22}, which then leads to the conclusion that magnetic order develops at an arbitrarily small $U$ in the $R \ra 0$ limit. 

We may estimate the magnitude of the antiferromagnetic order parameter, which corresponds to the density of states in the original dirty Weyl semimetal, 
using the following argument. 
Since the mass gap for the spinons arises due to monopole fluctuations in the compact QED$_3$, it should be of the same order of magnitude as the photon gap,
which comes from the same source~\cite{Fiebig90}. To estimate the photon mass, we perform the standard duality transformation of the compact QED$_3$ without 
matter~\cite{Polyakov_book,Herbut_book}.

We start from the partition function of the lattice version of QED$_3$ without fermions
\beq
\label{eq:44}
Z = \int_0^{2 \pi} \prod_{i \mu} d a_{i \mu} e^{K \sum \cos(\epsilon_{\mu \nu \lambda} \Delta_{\nu} a_{i \lambda})}, 
\eeq
where $K \sim t/U$ (here we will take hopping matrix elements to be equal to $t$ in all directions for simplicity) and the sum in the exponential is over plaquettes 
of the square lattice, on which the discrete curl $\epsilon_{\mu \nu \lambda} \Delta_{\nu} a_{i \lambda}$ is defined. 
Expanding the exponential, which is a periodic function of its argument, in Fourier series, we obtain the Villain (periodic Gaussian) form
\beq
\label{eq:45}
Z \approx \int_0^{2 \pi} \prod_{i \mu} d a_{i \mu} \sum_{h} e^{- \frac{1}{2 K} \sum_{j \mu} h_{j \mu}^2 + i \sum_{j \mu} h_{j \mu} \epsilon_{\mu \nu \lambda} \Delta_{\nu} 
a_{i \lambda}}, 
\eeq
where $h_{j \mu}$ are integer electromagnetic field variables, defined on the bonds of the dual lattice, normal to the plaquettes, on which the curl of $a$ is evaluated. 
Integrating over $a_{i \mu}$ gives
\beq
\label{eq:46}
\epsilon_{\mu \nu \lambda} \Delta_{\nu} h_{j \lambda} = 0, 
\eeq
which may be viewed as discrete Maxwell's equations. 
These may be solved as
\beq
\label{eq:47}
h_{j \mu} = \Delta_{\mu} n_j, 
\eeq
where $n_j$ are integer variables, defined on dual lattice sites. 
This gives the discrete Gaussian model
\beq
\label{eq:48}
Z = \prod_{i \mu} \sum_{n_{i \mu}} e^{- \frac{1}{2 K} \sum_{i \mu} (\Delta_{\mu} n_i)^2}. 
\eeq
Using Poisson summation formula, this may be written as
\beq
\label{eq:49}
Z = \prod_{i \mu} \int_{-\infty}^{\infty} d \varphi_i \sum_{n_{i \mu}} e^{- \frac{1}{8 \pi^2 K} \sum_{i \mu} (\Delta_{\mu} \varphi_i)^2 - i \varphi_i n_i}, 
\eeq
Integrating over $\varphi$ produces Coulomb interaction between $n_i$, which demonstrates that their physical meaning is monopole charges, 
while $\varphi_i$ is the corresponding ``electrostatic potential". 
We now imagine coarse-graining Eq.~\eqref{eq:49} by integrating high-energy modes of $\varphi$. 
Most importantly, this procedure will introduce an extra, self-interaction, term for the monopoles
\beq
\label{eq:50}
Z = \int_{-\infty}^{\infty} \prod_{i \mu} d \varphi_i \sum_{n_{i \mu}} e^{- \frac{1}{8 \pi^2 K} \sum_{i \mu} (\Delta_{\mu} \varphi_i)^2 - i \varphi_i n_i + \ln (y) \sum_i n_i^2}, 
\eeq
where 
\beq
\label{eq:51}
\ln(y) =  - \int \frac{d^3 q}{(2 \pi)^3} G(\bq) \sim - K \int_0^{\pi} \frac{d^3 q}{(2 \pi)^3} \frac{1}{q^2} = - \frac{C t}{U}.
\eeq
Here $y$ is the monopole fugacity, $G(\bq)$ is the Green's function of the ``electrostatic potential" $\varphi$ and $C$ is a numerical constant, whose 
precise value is unimportant. 
The fugacity is small in the $t/U \gg 1$ (i.e. weak disorder) limit, which we are interested in, corresponding to a dilute gas of monopoles. 
Expanding Eq.~\eqref{eq:50} in Taylor series with respect to $y$, we obtain the three-dimensional sine-Gordon model
\beq
\label{eq:52}
Z = \int_{-\infty}^{\infty} \prod_i d \varphi_i  e^{- \frac{1}{8 \pi^2 K} \sum_{i \mu} (\Delta_{\mu} \varphi_i)^2 + 2 y \sum_i \cos(\varphi_i)}.
\eeq
Unlike the same model in two dimensions, it is always in the gapped phase with a gap $\sim y$. 
This gap is the photon mass. 

We may now argue that the fermion mass, which arises entirely due to the monopole proliferation, must be proportional to the photon mass, with an $O(1)$ 
numerical coefficient, since both arise from the same source. Such a direct correlation between the monopole density and the chiral symmetry breaking order parameter has indeed been observed in numerical studies of compact QED$_3$ with matter~\cite{Fiebig90}.
This gives
\beq
\label{eq:53}
m \sim  e^{- \frac{C t}{U}}.
\eeq
As discussed above, the staggered magnetization $m$ has the meaning of the density of states at Fermi energy in the original 3D Weyl semimetal problem. 
This exponential dependence of the disorder-induced density of states on the variance of the disorder potential is consistent with previous work~\cite{Nandkishore14,Pixley16}.
Given this agreement, we can claim that the nonperturbative effect of rare regions, described in Refs.~\cite{Nandkishore14,Pixley16}, corresponds 
to (also nonperturbative) Mott physics within our mapping. 
\section{Nonlinear sigma model and absence of localization}
\label{sec:4}
We would now like to rederive the results of Ref.~\cite{Yi24} on the NLSM theory and absence of localization in Weyl semimetals,
using the coupled Hubbard chain picture, developed above. 
This will provide a more intuitive complementary picture of the lack of localization in magnetic Weyl semimetals. 

The absence of localization in this language is straightforward to understand. 
An Anderson insulator would correspond to a fully gapped state with no broken continuous symmetries, which result in gapless Goldstone (diffusion) modes. 
This requires gapping the pair of $2R$-degenerate 2D Dirac fermions without breaking symmetries.
This, however, is impossible, unless the nodes are moved to either the edge or the center of the BZ, which corresponds to a quantum Hall or an ordinary 
insulator states. The reason is that a state with a pair of 2D Dirac nodes, separated in momentum space, is topologically nontrivial. 

\begin{figure}[t]
\includegraphics[width=\columnwidth]{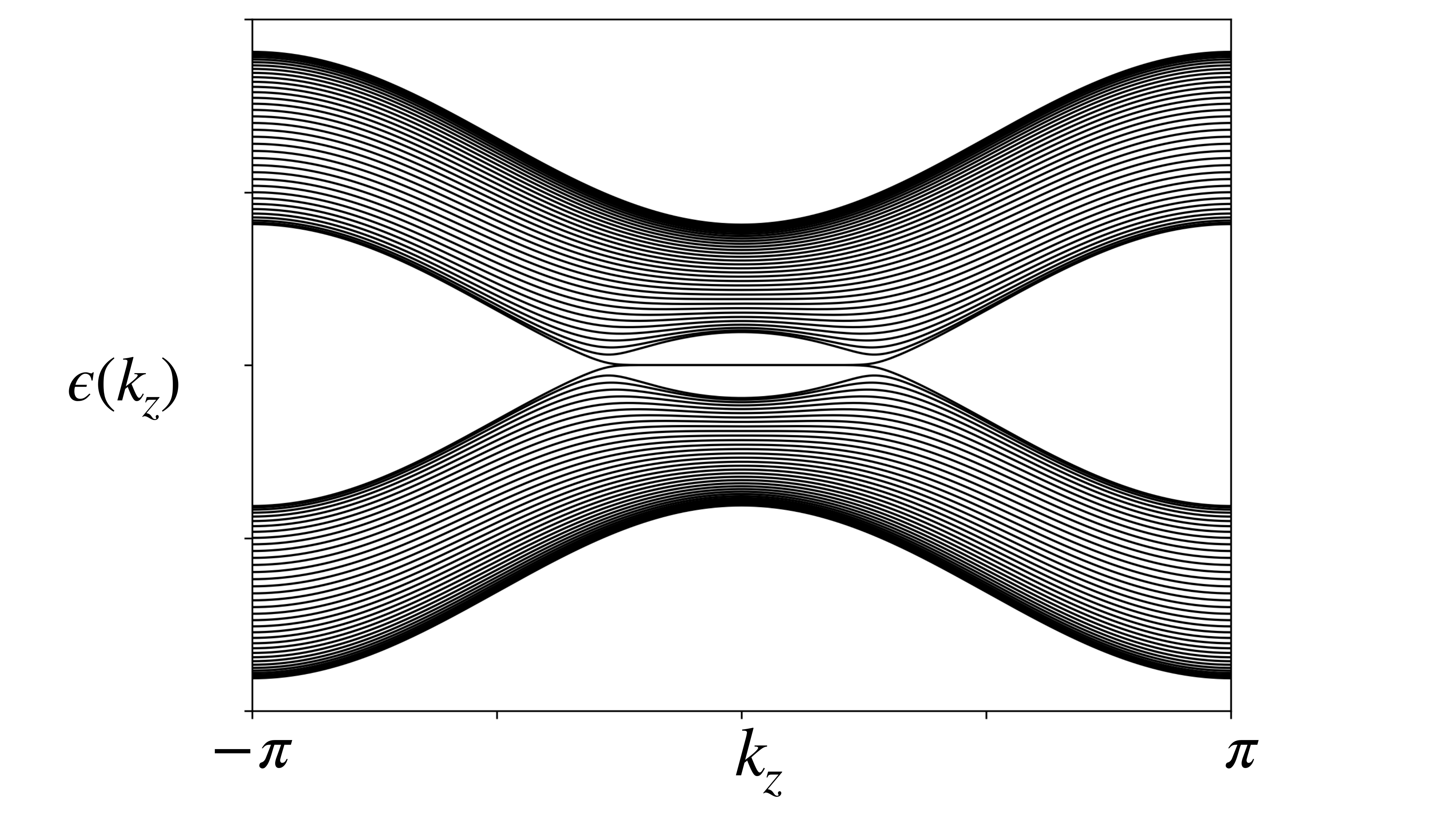}
\caption{Eigenstate spectrum of the kinetic energy part of the Hubbard array Hamiltonian Eq.~\eqref{eq:20} for a sample of finite size in the $x$-direction. 
Dispersionless edge states connect projections of the 2D Dirac nodes onto the edge BZ.}
\label{fig:2}
\end{figure}

To see this explicitly, it is convenient to consider a single replica. 
Let us take a sample that is finite in the $x$-direction. 
The kinetic energy part of the Hubbard Hamiltonian Eq.~\eqref{eq:20} may be written as 
\beq
\label{eq:54}
H = - i t \sigma^y \partial_x - 2 [t' \cos(k_z) + \delta t] \sigma^x = - i t \sigma^y \partial_x - m(k_z) \sigma^x. 
\eeq
This has the form of an array of 1D SSH chains~\cite{SSH} at every value of $k_z$. The ``mass" $m(k_z)$ changes sign at the 
locations of the Dirac nodes, signifying a transition from topological (with edge states) to ordinary (no edge states) phases of the SSH chain. 
A standard calculation gives the localized edge state
\beq
\label{eq:55}
\Psi(k_z,x) = e^{-\frac{1}{t} \int_0^x d x' m(k_z, x')} |\sigma^z = -1 \rangle. 
\eeq
This solution is dispersionless along the $z$-direction and exists for values of $k_z$ in between the locations of the Dirac nodes, see Fig.~\ref{fig:2}. 
The fact that the Dirac nodes are connected by this topological dispersionless edge state makes it clear that the nodes can not be gapped out without 
breaking the mirror symmetry Eq.~\eqref{eq:26.5}, which forces the edge state to appear at zero energy. 

Due to double retarded/advanced pseudospin degeneracy, at half-filling this state has zero charge and a pseudospin-$1/2$ per each value of $k_z$, if we apply 
the well-known counting of the edge mode quantum numbers in SSH chains~\cite{SSH}.
The existence of this charge-neutral spin-$1/2$ edge state may be expressed as the following Wess-Zumino (WZ) term in the imaginary time action of the edge 
\beq
\label{eq:56}
S_{WZ} = \frac{i \lambda}{2} \int d \tau d z [1 - \cos(\theta)] \partial_{\tau} \varphi, 
\eeq
where $\lambda = 2 Q/ 2 \pi$ is the ratio of the separation between the Dirac nodes to the reciprocal lattice vector (lattice constant is unity throughout) 
and $\bn = [\sin(\theta) \cos(\varphi), \sin(\theta) \sin(\varphi), \cos(\theta)]$ is the pseudospin-$1/2$ coherent state label. 
This edge state represents the chiral Fermi arc surface state of the 3D Weyl semimetal. Indeed, recalling that the imaginary time $\tau$ is simply the $y$-coordinate 
in the 3D Weyl problem, $\partial_{\tau} \varphi$ precisely represents chiral modes that propagate in the $y$-direction. 

Switching to a differential form notation and using the well-known expression for the area two-form on a unit sphere
\beq
\label{eq:57}
d [(1 - \cos(\theta)) d \varphi] = \sin(\theta) d \theta \wedge d \varphi = \frac{1}{2} \bn \cdot (d \bn \wedge d \bn), 
\eeq
this may be rewritten as the following bulk topological term
\beq
\label{eq:58}
S_{NLSM} = \frac{i \lambda}{4} \int \bn \cdot (e^z \wedge d \bn \wedge d \bn).
\eeq
Here $e^z$ is the ``translation gauge field"~\cite{Volovik19,Wang21,Song21,Else_QC,Hughes24},
which is defined as the gradient of a phase function $\theta^z(\br, \tau)$, 
\beq
\label{eq:59}
e^z_{\mu} = \frac{1}{2 \pi} \partial_{\mu} \theta^z.
\eeq
The physical meaning of $\theta^z(\br, \tau)$ is that the solutions of 
\beq
\label{eq:60}
\theta^z(\br, \tau) = 2 \pi n, 
\eeq
where $n \in \mathbb{Z}$, define the family of crystal planes, perpendicular to the $z$-direction. 

Eq.~\eqref{eq:58} may be rewritten using a more common notation by introducing a matrix 
\beq
\label{eq:61}
Q_{1 1} = \bn \cdot \btau, \,\, Q_{a b} = \delta_{a b} \tau^z, \,\, a > 1, 
\eeq
where $a, b$ are replica indices. 
Then it takes the form 
\beq
\label{eq:62}
S_{NLSM} = \frac{\lambda}{8} \int \textrm{Tr}(Q e^z \wedge d Q \wedge d Q).
\eeq
This is precisely the topological term, derived in Refs.~\cite{Wang15,Altland16,Yi24} using different arguments. 
As was shown in Ref.~\cite{Yi24}, any noninteger value of $\lambda$ is incompatible with localization, which in the present 
context has a simple explanation, namely the impossibility of gapping a pair of 2D Dirac fermions without breaking the retarded/advanced 
pseudospin $SU(2)$ symmetry. 

\section{Discussion and conclusions}
\label{sec:5}
In this paper we have used an analogy between the 3D magnetic Weyl semimetal and the 2D quantum Hall plateau transition to address 
the question of the effect of disorder on the Weyl nodes. 
The analogy leads to a 3D generalization of the Chalker-Coddington network model, which in turn can be mapped onto an array of coupled replicated Hubbard chains at 
half-filling in the zero replica $R \ra 0$ limit. The variance of the disorder potential, assuming a gaussian distribution, maps directly onto the Hubbard $U$. 
We have demonstrated that this array of Hubbard chains is in the Mott insulator phase at any nonzero $U$ in the limit $R \ra 0$ and magnetic order in the retarded/advanced pseudospin space is always present. 
This translates into a nonzero density of states at the Weyl nodes and diffusive transport at long times and long distances 
at arbitrarily weak disorder, which is consistent with previous work~\cite{Nandkishore14,Pixley15,Pixley16,Altland18}.

One may also discuss the strong disorder regime using the same language. An important question in this regime is whether the Weyl metal state at weak disorder gives way to an Anderson insulator. 
In Ref.~\cite{Yi24} we presented arguments, based on the matrix NLSM, as well as a decorated domain wall construction, that localization in this system is impossible, as long as disorder does not destroy the 3D quantum anomalous Hall insulator state. 
Here we have revisited this problem from the viewpoint of the coupled Hubbard chain mapping. 
In this language, localization in a 3D Weyl metal may be recast as the problem of symmetric mass generation for 2D Dirac fermions. 
It was demonstrated to be impossible due to a nontrivial topology of this system, namely the presence of edge states, connecting 
the Dirac node projections onto the 1D edge BZ, and a corresponding topological term. 
We have related this to the earlier arguments, based on the matrix NLSM. 

An interesting extension of the present work would be to directly explore the large-$U$ limit of the Hubbard array, in which it maps onto an $SU(2 R)$ spin model
with antiferromagnetic spin-spin interactions $J \sim t^2/U$, following the same pattern as the tunneling amplitudes in Fig.~\ref{fig:1}. 
This relatively simple spin model must exhibit a phase with highly nontrivial 1D edge states, corresponding to the chiral Fermi arcs of the disordered Weyl metal. 
\begin{acknowledgments}
We thank Igor Herbut, Sung-Sik Lee and Chong Wang for useful discussions.  
Financial support was provided by the Natural Sciences and Engineering Research Council (NSERC) of Canada.
AAB was supported by Center for Advancement of Topological Semimetals, an Energy Frontier Research Center funded by the U.S. Department of Energy Office of Science, Office of Basic Energy Sciences, through the Ames Laboratory under contract DE-AC02-07CH11358. 
Research at Perimeter Institute is supported in part by the Government of Canada through the Department of Innovation, Science and Economic Development and by the Province of Ontario through the Ministry of Economic Development, Job Creation and Trade.
\end{acknowledgments}
\bibliography{references}
\end{document}